\documentclass[]{spie}  

\usepackage{amsmath,amsfonts,amssymb}
\usepackage{graphicx}
\usepackage{subfigure}
\usepackage{multirow}
\usepackage[colorlinks=true, allcolors=blue]{hyperref}

\title{Project Management for Ground-based Telescope Array Development}

\author[a,b]{Ji Hoon Kim}
\author[a, b]{Myungshin Im}
\author[a, b]{Hyung Mok Lee}
\author[a,b]{Seo-Won Chang}
\affil[a]{Astronomy Research Center, Seoul National University, Seoul, Republic of Korea}
\affil[b]{Astronomy Progrma, Department of Physics and Astronomy, Seoul National University, Seoul, Republic of Korea}

\authorinfo{Further author information: (Send correspondence to Ji Hoon Kim.)\\Ji Hoon Kim: jhkim.astrosnu@gmail.com,
Myungshin Im: myungshin.im@gmail.com}

\begin{document} 
\maketitle

\begin{abstract}
Center for the Gravitational-Wave Universe at Seoul National University has been operating its main observational facility, the 7-Dimensional Telescope (7DT) since October 2023.
Located at El Sauce Observatory in Chilean Rio Hurtado Valley, 7DT consists of 20 50-cm telescopes equipped with 40 medium-band filters of 25 nm full width at half maximum along with a CMOS camera of 61 megapixels.
7DT produces about 1 TB per night of spectral mapping image data including calibration, and the byproduct of the data reduction pipeline once our planned three layered surveys (Reference Imaging Survey, Wide Field Survey, and Intensive Monitoring Survey) start in 2024.
We are expecting to generate 1 PB per year by combining raw data, reduced data, and data products (e.g. calibrated stacked images, spectral cubes, and object catalogs).
To incorporate this huge amount of data, we now have a data storage for 1 PB which we will increment by 1 PB per year.
We also have a high-performance computation facility that is equipped with 2 NVIDIA A100 GPU cards since we plan to carry out real-time data reduction and analysis for follow-up observation data of gravitational wave events.
To incorporate this, we established a 400 Mbps network connection between the facilities in Korea and Chile.
Taking advantage of the high-performance network, we have been carrying out fully remote operations since October 2023. 
In this talk, we present details of designing, planning, and executing the ground-based telescope facility project, especially within low-budget academic environments.
While we cover as much ground as possible, we will emphasize human resource management, project risk management, and financial contingency management.

\end{abstract}


\section{INTRODUCTION}
\label{sec:intro}  

The discovery of the gravitational-wave (GW) electromagnetic counterpart, GW170817, ignited the field of multimessenger astronomy (MMA)\cite{2017ApJ...848L..12A}.
However, it remains the only confirmed case. 
Kilonovae (KNe), resulting from compact binary coalescence events involving at least one neutron star (NS), are rare and elusive.

Detecting KNe is challenging due to their low luminosities, large localization areas, and numerous transients and spurious objects within these areas.
KNe are typically about 3 magnitudes fainter than Type Ia supernovae (SNe) at the peak and their brightness declines rapidly, roughly 0.5 magnitude per day in the optical wavelengths.
During the O3 run of the GW detector network, LVK, including the Advanced Laser Interferometer Gravitational-wave Observatory (LIGO)\cite{2015CQGra..32g4001L}, Advanced Virgo\cite{2015CQGra..32b4001A}, and Kamioka Gravitational Wave Detector (KAGRA)\cite{2021PTEP.2021eA101A}, localization areas ranged from hundreds to thousands of square degrees\cite{2020LRR....23....3A}.

Despite the O4 run starting on May 24, 2023, and expectations for significant improvements, precise localization remains challenging.
These vast areas are difficult to cover with optical telescopes and yield numerous spurious objects, including other transients and detector artifacts.

To address these challenges, optical follow-up facilities must rapidly cover extensive areas with large field-of-view (FoV) telescopes and flexible operations.
Distinguishing true GW electromagnetic counterparts from spurious objects requires rigorous validation.
While ruling out detector artifacts is straightforward, separating KNe from SNe necessitates spectroscopic classification, even with low spectral resolution.

Center for the Gravitational-wave Universe at Seoul National University, a research center whose main scientific endeavors focus on GW physics and MMA established an observational facility, the 7-Dimensional Telescope (7DT).
A multi-telescope array with 20 telescopes, 7DT utilizes 20 50-cm commercial off-the-shelf (COTS) modified Cassegrain telescopes with a focal ratio of f/3.
Each unit of 7DT is equipped with a CMOS camera whose detector has 9576 by 6388 pixels of 3.76 $\mu m$.
At f/3, the camera provides an FoV of 1.33 by 0.89 degrees with a projected pixel size of 0.5 by 0.5 arcseconds.

7DT uses 40 medium-band filters with a full-width half maximum of 25 nm that cover the wavelength range from 400 nm through 900 nm.
The gaps between these medium-band filters are 12.5 nm.
Currently, 12 telescopes are deployed at the site and are operational with 2 sets of 20 medium-band filters whose central wavelengths range from 400 nm to 875 nm with a gap of 25 nm.

7DT is located in the Chilean Rio Hurtado Valley and shares the sky conditions with Cerro Tololo Inter-American Observatory, Gemini South Telescope, the Southern Astrophysical Research Telescope, and the Vera C. Rubin Observatory.

At our home institution, we set up a data storage system of 1PB to accommodate a large data influx, up to 1 TB per night.
We will increment the volume of the data storage by 1 PB per year up to 5 PB.
To carry out real-time data reduction and spectral classification, we set up a data processing system equipped with two NVIDIA A100 GPUs.

The observing facility in Chile and the computing facilities at Seoul National University are connected by Korea Research Environment Open Network (KREONET) which is operated and maintained by Korea Institute of Science and Technology Information (KISTI).
Overall, 7DT is designed to provide low-resolution spectroscopy with a large Fov and is capable of detecting candidates of GW electromagnetic counterparts with its flexible operation capability and real-time data reduction and analysis pipeline.

In this work, we present the practical aspects of project management involved in developing a ground-based telescope array development at a remote site, utilizing COTS instruments within academic settings.

\begin{figure} [ht]
\begin{center}
\begin{tabular}{cc} 
\subfigure[]{\includegraphics[width=0.5\textwidth]{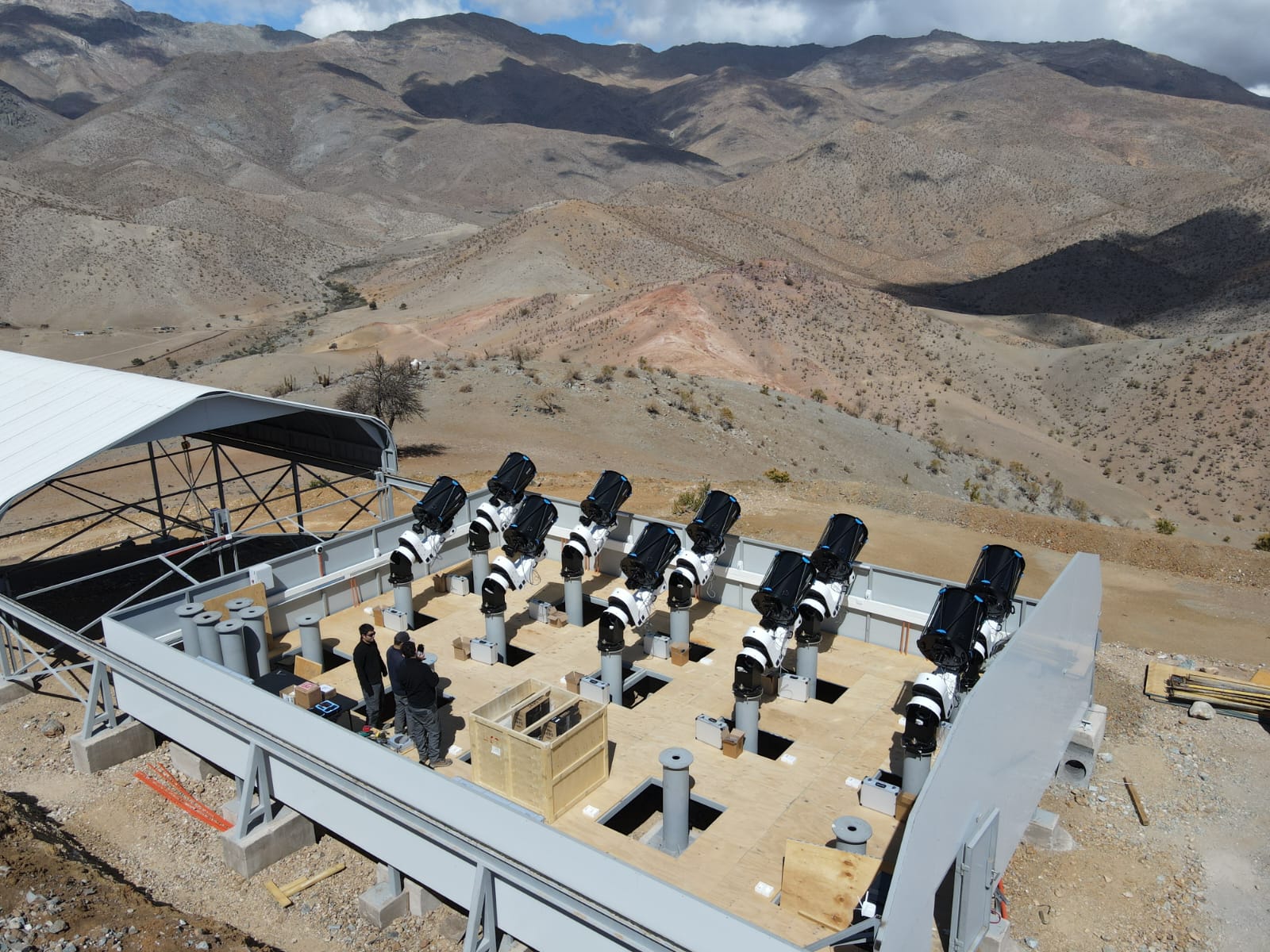}}
\subfigure[]{\includegraphics[width=0.48\textwidth]{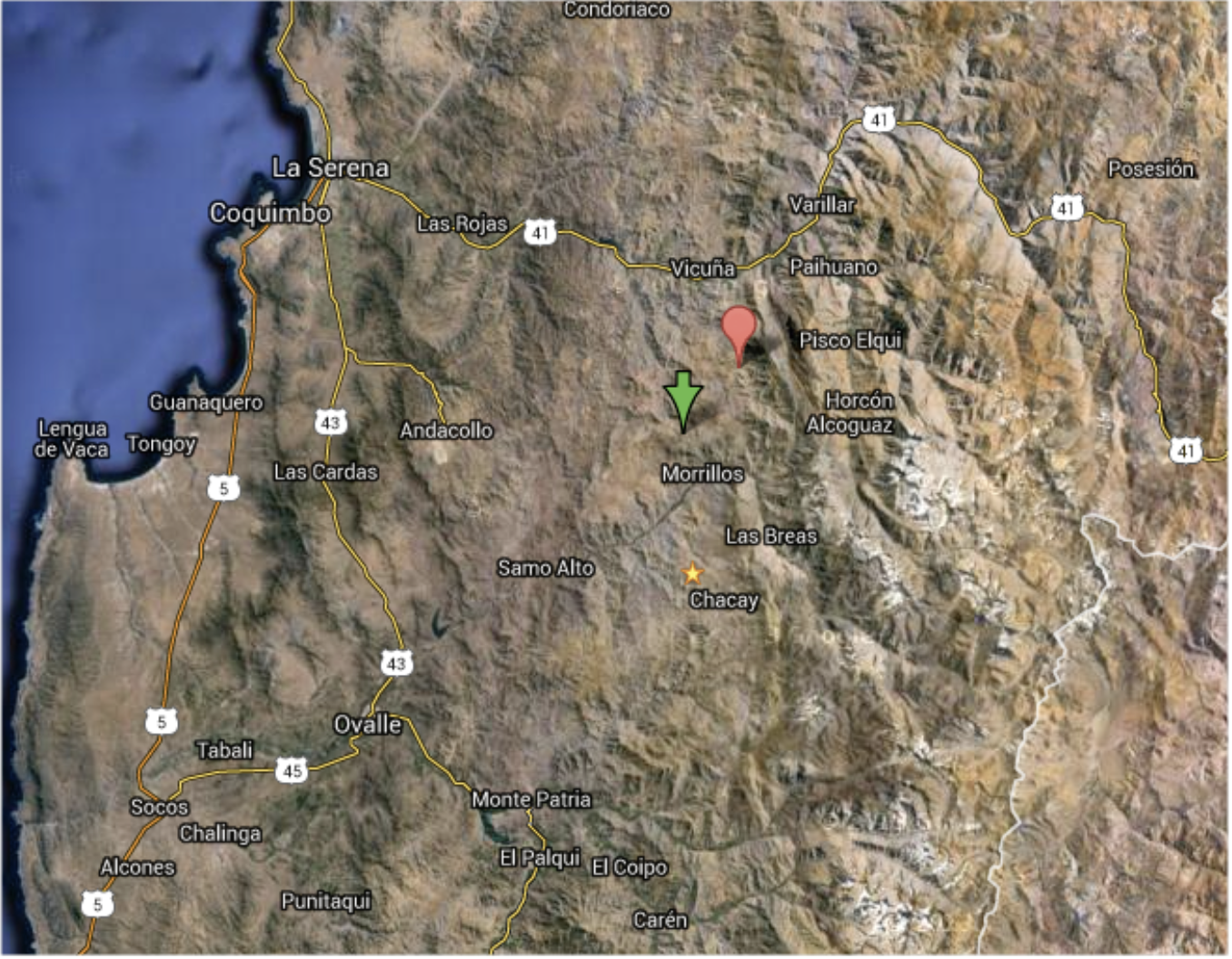}}

\end{tabular}
\end{center}

\caption 
{\label{fig:site} 
(a) A bird's eye view of 7DT at El Sauce Observatory in the Chilean Rio Hurtado Valley.
The infrastructure and maintenance of the site are provided by ObsTech. The picture shows 10 units of DeltaRho 500 installed as of September 2023.
(b) The location of the El Sauce Observatory is indicated with a yellow star. Green and Orange arrows show the respective locations of CTIO and Cerro Pachon.}

\end{figure} 

\section{Background}
\label{sec:bg}

\subsection{Budget}

The main funding resource for the Korean scientific community is the National Research Foundation of Korea (NRF) which is a subsidiary organization under the Ministry of Science and Information and Communication Technology.
NRF supports various research projects with different funding schemes for individual researchers and research groups.
Center for the Gravitational-wave Universe is supported by the funding project called `National Convergence Research of Scientific Challenges (NCRSC)' which is the biggest funding class from NRF.
Between the two subclasses of NCRSC, our center belongs to `National Science Challenge Initiatives' that supports 1.75 million USD annually for up to five years.
Last year, there was an announcement that the research budget not only for NRF, but also for all the Korean government funding agencies would be reduced.
The budget for the fiscal year of 2024 is reduced to 75\% while it is unclear whether the budget for the fiscal year of 2025 will be restored to its original plan.

\subsection{Human Resource}

Although Center for the Gravitational-wave Universe is composed of members from several research institutes, namely Seoul National University, Ewah Womans University, POSTECH (Pohang University of Science and Technology), and Korea Astronomy and Space Science Institute, the project activities involved the development of 7DT is mainly carried out by the members at Seoul National University.
Seoul National University is the oldest and the biggest national university in the Republic of Korea.
Thus, the Department of Physics and Astronomy at Seoul National University to which the members of the 7DT project belong, is one of the biggest academic organizations in Korea.
However, the project's human resources are still relatively small compared to international competitions.
Overall, we have three doctorate researchers, and 12 graduate students on top of one faculty member.

\subsection{Location}

7DT is hosted at El Sauce Observatory in Rio Hurtado Valley whose infrastructure and maintenance are provided by a Chilean telescope company, ObsTech.
The location of the site is 30$^\circ$28$'$16$''$ S and 70$^\circ$45$'$47$''$ W.
This site is near Cerro Tololo Inter-American Observatory, Gemini South Telescope, the Southern Astrophysical Research Telescope, and the Vera C. Rubin Observatory, sharing excellent sky conditions with these facilities.
The site typically experiences a seeing of 1.5$''$, with an annual average of 300 clear nights.
The mean sky brightness at the zenith, measured by a sky quality meter, is 27.97 mag.
Meanwhile, the production facility of the telescope manufacturer is located in Adrian, Michigan.

\subsection{COVID-19}

7DT utilizes 20 telescopes that adopt a corrected Cassegrain optical design.
The design implements a lens set that consists of three lenses.
During the early production phase of 7DT's telescopes, which was in early 2022, the supply of the lens material was limited due to the lingering effects of COVID-19.
Therefore, the production of the lens sets, consequently, the telescope units were delayed.

Not only were the instruments for the observing facility impacted, but also the computing facilities.
Supply chain disruptions, factory shutdowns, and increased demands for electronics strained semiconductor production.
Shortages of semiconductor production led to delays in the production of data storage, network essentials, CPU, and GPU while increasing prices.

\section{SWOT Analysis}
\label{sec:swot}

In this section, we analyze our project and its progress based on the SWOT analysis method.
Standing for Strengths, Weaknesses, Opportunities, and Threats, the SWOT analysis is a strategic planning and management technique.

\subsection{Strength}

One of the key strengths highlighted by the SWOT analysis is the cost-effectiveness achieved through the use of COTS instruments.
By integrating COTS instruments into projects, significant cost savings can be realized.
Developing customized instruments requires a research budget that is at least fourfold and is time-consuming.
The economies of scale enjoyed by COTS manufacturers, who produce these instruments in large quantities, translate into lower unit costs.

Another notable strength is the efficiency gained through outsourcing instrument production and facility management.
By partnering with specialized vendors and service providers, projects can benefit from their expertise and experience, leading to quicker implementation.
These external partners bring a wealth of field knowledge and technical proficiency, ensuring that instruments are produced to consistent standards and that facilities are managed optimally.
This outsourcing enables project teams to focus on core activities, such as research and development, without being hindered by the complexities of instrument production and facility upkeep.
The quick implementation facilitated by these partnerships also accelerates project timelines, allowing for faster realization of project goals.

Finally, the combination of cost savings from using COTS instruments and the operational efficiencies from outsourcing results in more efficient budget allocation.
The financial resources saved can be redirected to other critical areas of the project, such as research and educational activities.
This flexibility in budget management ensures that funds are utilized where they are most needed, enhancing the overall effectiveness and success of the project.
Efficient budget allocation also provides a buffer for unforeseen expenses, contributing to the project's resilience and long-term sustainability.

\subsection{Weakness}

There are several weaknesses that need to be addressed for the successful execution of the project. 

The most significant challenge stemming from the use of COTS instruments is mismatched specification and performance issues.
These COTS instruments are produced by manufacturers with limitied familiarity with the specific requirements of scientific research.
Consequently, commercial producers may not fully grasp the stringent standards and precise needs of scientific instrumentation.
As a result, the instruments may fall short in meeting the specialized demands of scientific projects, potentially compromising research quality and outcomes.

Another inherent weakness of relying on COTS products is the limited opportunity for customization and performance optimization.
COTS instruments are designed for broad, general use and may not offer the flexibility needed to tailor them to the unique needs of individual projects.
This limitation can hamper the ability to accomplish optimal performance and adapt instruments to specific research objectives.
The lack of customization options may also restrict the scope of scientific inquiries and the ability to explore novel or niche research areas effectively.

Additionally, we often face communication delays due to time zone differences that present a significant operational challenge.
When collaborating with international partners or suppliers, these time differences can lead to lag in responses, slowing down decision-making processes and execution timelines.
Delays in communication can cascade into broader project delays, affecting milestones and overall project progress.

The remote location of the observing facility poses another weakness, particularly in terms of providing hands-on experiences for student researchers.
These remote sites can limit access for educational purposes, depriving students of valuable practical training and experiential learning opportunities.
The distance barrier can hinder student engagement and reduce the effectiveness of educational programs associated with the project.

Lastly, the shortage of manpower is a critical issue that can overburden existing project members.
With limited personnel available to manage the various aspects of the project, project members can become overwhelmed with responsibilities, leading to decreased productivity and increased stress levels.
Overburdened project members may struggle to maintain high standards of work, which can affect the overall quality and efficiency of the project.

\subsection{Opportunity}

There are several opportunities that can be leveraged to enhance the values of the project's heritage in terms of both scientific and educational goals.

One significant opportunity is the early responsibility and rapid growth offered to students involved in the project.
By engaging students in meaningful roles from the outset, the project provides them with valuable hands-on experience and fosters a steep learning curve.
This early involvement not only accelerates their professional development but also cultivates a strong sense of ownership and commitment to the project.
As students gain practical skills and confidence, they become more capable and resourceful contributors to the research community.

Another notable opportunity arises from the facility's location, which boasts some of the best weather conditions for astronomical observations.
Clear skies and favorable atmospheric conditions enhance the quality and consistency of data acquisition, making our planned survey capable of providing data for cutting-edge astronomical reseasrch.
This geographic advantage can attract additional interest and investment in the facility, further elevating its status within the scientific community.

The location of the facility, while presenting some logistical challenges, also offers a unique opportunity to increase the project's visibility within the astronomical research community.
One of the best sites for astronomical observatories, our facility is surrounded by a few of the best and biggest observatories in the world.
Being unique in terms of its instrument configuration and management, its visibility is elevated.

Furthermore, the hightened visibility of the project opens up numerous opportunities for research collaborations.
As the project gains recognition, it can attract partnerships with other institutions, researchers, and organizations.
These collaborations can lead to a richer exchange of ideas, resources, and expertise, fostering innovation and expanding the scope and impact of the research conducted.
Collaborative efforts can also enhance funding prospects, as joint projects often have a stronger appeal to grant-making bodies and sponsors.

\subsection{Threat}
The following are several critical threats that pose challenges to the projects.

A major threat stems from the ongoing budget crisis.
This financial strain has resulted in a 25\% reduction in the budget allocated to our project for the current year and the following.
Such a substantial cut in funding can severely impact various aspects of the project, from resource allocation to research activities, potentially compromising the project's overall progress and outcomes.
Most critically, we are considering downsizing the number of telescope units to 16 from 20, unless we secure another funding to support the project.

Another significant threat is the adverse fluctuation in the exchange rate between the Korean Won (KRW) and the US Dollar (USD).
Since the project commenced in the year of 2021, the exchange rate has increased by more than 25\%, effectively raising the cost of imported materials and equipment.
This currency volatility exacerbates budgetary constraints, making it more challenging to procure necessary resources and maintain financial stability.
The increased costs due to unfavorable exchange rates can strain the project's budget further, limiting its capacity to achieve its objectives.

The manufacturing of instruments experienced severe delays during the COVID-19 pandemic, which affected the early phase of the project.
The disruptions in global supply chains and manufacturing processes led to significant setbacks in the production and delivery of essential instruments.
These delays not only hinder the project's progress but also reduce the time available for critical testing and calibration activities.
One consoling aspect of these delays that the science operation of LVK was also delayed and its detection sensitivity is not par with the expectation.

Finally, the accumulated delays result in lost opportunities and intensify competition within the scientific community.
As the project timeline extends, other research groups may advance more rapidly, capturing opportunities that our project could have otherwise seized.
The increased competition for limited resources, recognition, and collaboration opportunities can further challenge the project's ability to achieve its goals and maintain its competitive edge.
The pressure to keep up with or surpass competing projects adds layer of stress and urgency to the project's execution.

\begin{table}[ht]
\caption{SWOT Anaysis} 
\label{tab:SWOT}
\begin{center}       
\begin{tabular}{|c|c|c|} 
\hline
\rule[-1ex]{0pt}{3.5ex}   & Helpful  & Harmful  \\
\hline
\rule[-1ex]{0pt}{3.5ex} \multirow{5}{*}{Internal} & \multicolumn{1}{l|}{\bf Strength} & \multicolumn{1}{l|}{\bf Weakness}\\
\rule[-1ex]{0pt}{3.5ex}   & Cost Effective & Limited scientific knowledge   \\
\rule[-1ex]{0pt}{3.5ex}   & Quick Implementation & No customization and performance optimization   \\
\rule[-1ex]{0pt}{3.5ex}   & Professional Expertise & Less hands-on experience for students   \\
\rule[-1ex]{0pt}{3.5ex}   & Efficient budget allocation & Overburdening personnel   \\
\hline
\rule[-1ex]{0pt}{3.5ex} \multirow{5}{*}{External} & \multicolumn{1}{l|}{\bf Opprotunity} & \multicolumn{1}{l|}{\bf Threat}\\
\rule[-1ex]{0pt}{3.5ex}   & Students' opportunity & Unstable fiscal situation due to budget cut   \\
\rule[-1ex]{0pt}{3.5ex}   & Optimal sky conditions & Exchange rate fluctuation   \\
\rule[-1ex]{0pt}{3.5ex}   & Better visibility & Delays due to COVID-19  \\
\rule[-1ex]{0pt}{3.5ex}   & More external collaboration & Lost opportunity and more competition   \\

\hline 
\end{tabular}
\end{center}
\end{table}

\section{Conclusion}

In conclusion, the strengths derived from using COTS instruments, outsourcing production and management, and achieving efficient budget allocation collectively enhance the project's cost-effectiveness, operational efficiency, and financial flexibility.
These advantages position the project for successful implementation and sustainable growth.
The weaknesses identified include the limited understanding of scientific requirements by commercial producers, restricted customization and performance optimization of COTS products as well as communication delays due to time zone differences, the prohibitive nature of remote locations for hands-on student experiences, and the shortage of manpower leading to overburdened staff.
The opportunities for the project include the rapid growth of students thanks to early responsibility afforded to students, the excellent weather conditions for astronomical observations, the increased visibility of the project due to its remote location, and the potential for expanded research collaborations.
The threats identified are the budget crisis within the Korean science community, unfavorable exchange rate fluctuations, manufacturing delays caused by the COVID-19 pandemic, and the resulting lost opportunities and heightened competition.

For observing facility projects with small-aperture telescopes, It is imperative to utilize COTS instruments to mitigate development costs and adhere project timelines.
However, while outsourcing professional expertise in instrument development and facility maintenance offers numerous advantages, meticulous planning is essential for their success.
Additionally, the financial stability of such projects remains vulnerable to external factors like budget cuts, exchange rate fluctuations, and international affairs, necessitating the presence of a robust contingency plan.
Moreover, instrument projects within academic environments must harmonize research goals with educational objectives, ensuring a balanced approach that fosters both project advancement and student development.

\subsection{Recommnedations \& Resolution}

To navigate the complexities of observing facility projects effectively, it is essential to consider a range of factors that facilitate to accomplish goals of projects. 
Firstly, understanding the limitations of COTS instruments is essential, ensuring to align between project requirements and instrument capabilities.
Secondly, cultivating mutual understanding between scientific researchers and practitioners facilitates more efficient project execution, enhancing collaboration and synergy.
Thirdly, maintaining frequent communication between researchers and practitioners fosters transparency and coherence, allowing for timely adjustments and problem-solving while removing misunderstanding.
Additionally, adopting a flexible approach to planning, encompassing scheduling, budgeting, human resources, and scientific goals, enables adaptability in the face of evolving project needs.
Lastly, the efficient allocation of human resources across various project aspects, including data reduction and analysis, operations, database management, and administration, optimizes project efficiency and effectiveness.

\acknowledgments 

The 7-Dimensional Telescope is designed, built, and operated by Center for the Gravitational-wave Universe at Seoul National University which is supported by the National Research Foundation of Korea (NRF) grant, No. 2021M3F7A1084525 by the Korea government (MSIT).
We also acknowledge the support from the National Research Foundation of Korea (NRF) grant, No. 2020R1A2C3011091.
This work was supported by KREONET(Korea Research Environment Open NETwork)/KISTI(Korea Institute of Science and Technology Information).
J.H.K acknowledges the support from the Institute of Information \& Communications Technology Planning \& Evaluation (IITP) grant, No. RS-2021-II212068.
SWC acknowledges the support from the Basic Science Research Program through the NRF funded by the Ministry of Education (RS-2023-00245013).
\bibliography{SPIE_13099} 
\bibliographystyle{spiebib} 

\end{document}